\documentclass[twocolumn,prl,aps,superscriptaddress,floatfix]{revtex4}
\usepackage{graphicx}
\begin{document}

\title{Production of $e^+e^-$ Pairs Accompanied by Nuclear
Dissociation in Ultra-Peripheral Heavy Ion Collisions}

\affiliation{Argonne National Laboratory, Argonne, Illinois 60439}
\affiliation{University of Birmingham, Birmingham, United Kingdom}
\affiliation{Brookhaven National Laboratory, Upton, New York 11973}
\affiliation{California Institute of Technology, Pasedena, California 91125}
\affiliation{University of California, Berkeley, California 94720}
\affiliation{University of California, Davis, California 95616}
\affiliation{University of California, Los Angeles, California 90095}
\affiliation{Carnegie Mellon University, Pittsburgh, Pennsylvania 15213}
\affiliation{Creighton University, Omaha, Nebraska 68178}
\affiliation{Nuclear Physics Institute AS CR, 250 68 \v{R}e\v{z}/Prague, Czech Republic}
\affiliation{Laboratory for High Energy (JINR), Dubna, Russia}
\affiliation{Particle Physics Laboratory (JINR), Dubna, Russia}
\affiliation{University of Frankfurt, Frankfurt, Germany}
\affiliation{Insitute  of Physics, Bhubaneswar 751005, India}
\affiliation{Indian Institute of Technology, Mumbai, India}
\affiliation{Indiana University, Bloomington, Indiana 47408}
\affiliation{Institut de Recherches Subatomiques, Strasbourg, France}
\affiliation{University of Jammu, Jammu 180001, India}
\affiliation{Kent State University, Kent, Ohio 44242}
\affiliation{Lawrence Berkeley National Laboratory, Berkeley, California 94720}
\affiliation{Massachusetts Institute of Technology, Cambridge, MA 02139-4307}
\affiliation{Max-Planck-Institut f\"ur Physik, Munich, Germany}
\affiliation{Michigan State University, East Lansing, Michigan 48824}
\affiliation{Moscow Engineering Physics Institute, Moscow Russia}
\affiliation{City College of New York, New York City, New York 10031}
\affiliation{NIKHEF, Amsterdam, The Netherlands}
\affiliation{Ohio State University, Columbus, Ohio 43210}
\affiliation{Panjab University, Chandigarh 160014, India}
\affiliation{Pennsylvania State University, University Park, Pennsylvania 16802}
\affiliation{Institute of High Energy Physics, Protvino, Russia}
\affiliation{Purdue University, West Lafayette, Indiana 47907}
\affiliation{University of Rajasthan, Jaipur 302004, India}
\affiliation{Rice University, Houston, Texas 77251}
\affiliation{Universidade de Sao Paulo, Sao Paulo, Brazil}
\affiliation{University of Science \& Technology of China, Anhui 230027, China}
\affiliation{Shanghai Institute of Applied Physics, Shanghai 201800, P.R. China}
\affiliation{SUBATECH, Nantes, France}
\affiliation{Texas A\&M University, College Station, Texas 77843}
\affiliation{University of Texas, Austin, Texas 78712}
\affiliation{Tsinghua University, Beijing, P.R. China}
\affiliation{Valparaiso University, Valparaiso, Indiana 46383}
\affiliation{Variable Energy Cyclotron Centre, Kolkata 700064, India}
\affiliation{Warsaw University of Technology, Warsaw, Poland}
\affiliation{University of Washington, Seattle, Washington 98195}
\affiliation{Wayne State University, Detroit, Michigan 48201}
\affiliation{Institute of Particle Physics, CCNU (HZNU), Wuhan, 430079 China}
\affiliation{Yale University, New Haven, Connecticut 06520}
\affiliation{University of Zagreb, Zagreb, HR-10002, Croatia}

\author{J.~Adams}\affiliation{University of Birmingham, Birmingham, United Kingdom}
\author{M.M.~Aggarwal}\affiliation{Panjab University, Chandigarh 160014, India}
\author{Z.~Ahammed}\affiliation{Variable Energy Cyclotron Centre, Kolkata 700064, India}
\author{J.~Amonett}\affiliation{Kent State University, Kent, Ohio 44242}
\author{B.D.~Anderson}\affiliation{Kent State University, Kent, Ohio 44242}
\author{D.~Arkhipkin}\affiliation{Particle Physics Laboratory (JINR), Dubna, Russia}
\author{G.S.~Averichev}\affiliation{Laboratory for High Energy (JINR), Dubna, Russia}
\author{Y.~Bai}\affiliation{NIKHEF, Amsterdam, The Netherlands}
\author{J.~Balewski}\affiliation{Indiana University, Bloomington, Indiana 47408}
\author{O.~Barannikova}\affiliation{Purdue University, West Lafayette, Indiana 47907}
% ...
\collaboration{STAR Collaboration}\homepage{www.star.bnl.gov}\noaffiliation

\begin{abstract}

We present the first data on $e^+e^-$ pair production accompanied by
nuclear breakup in ultra-peripheral gold-gold collisions at a center
of mass energy of 200 GeV per nucleon pair.  The nuclear breakup
requirement selects events at small impact parameters, where
higher-order diagrams for pair production should be enhanced.  We
compare the data with two calculations: one based on the equivalent
photon approximation, and the other using lowest-order quantum
electrodynamics (QED).  The data distributions agree with both
calculations, except that the pair transverse momentum spectrum
disagrees with the equivalent photon approach.  We set limits on
higher-order contributions to the cross section.

\end{abstract}


\maketitle

Electron-positron pairs are copiously produced by photon interactions
in the strong electromagnetic fields of fully stripped colliding heavy
nuclei (cf. Fig. 1); the field strength at the surface of the ions
reaches $10^{20}$ V/cm!  At a center of mass energy of $\sqrt{s_{NN}}
= 200$ GeV per nucleon pair, the production cross section is expected
to be 33,000 b, or 4,400 times the hadronic cross section
\cite{reviews,baurlum}.

The electromagnetic fields are strong enough, with coupling
$Z\alpha\approx 0.6$, ($Z$ is the nuclear charge and $\alpha \approx
1/137$ the fine-structure constant), that conventional perturbative
calculations of the process are questionable.  Many groups have
studied higher-order calculations of pair production.  Some early
coupled-channel calculations predicted huge (order-of-magnitude)
enhancements in the cross section \cite{rumrich} compared to 
lowest-order perturbative calculations.

Ivanov, Schiller and Serbo \cite{ivanov} followed the Bethe-Maximon
approach \cite{bethemaxmon}, and found that at RHIC, Coulomb
corrections to account for pair production in the electromagnetic
potential of the ions reduce the cross section 25\% below the
lowest-order result.  For high-energy real photons incident on a heavy
atom, these Coulomb corrections are independent of the photon energy
and depend only weakly on the pair mass \cite{bethemaxmon}.  However,
for intermediate-energy photons, there is a pair-mass
dependence, and also a difference between the $e^+$ and $e^-$ spectra
due to interference between different order terms \cite{lee2}.

In contrast, initial all-orders calculations based on solving the
Dirac equation exactly in the ultra-relativistic limit \cite{nonpert}
found results that match the lowest-order perturbative result
\cite{racah}.  However, improved all-orders calculations have
agreed with the Coulomb corrected calculation \cite{lee}.  These all-orders
calculations do not predict the kinematic distributions of the
produced pairs.

Any higher-order corrections should be the largest close to the
nuclei, where the photon densities are largest.  These high-density
regions have the largest overlap at small ion-ion impact parameters,
$b$.  Small-$b$ collisions can be selected by choosing events where
the nuclei undergo Coulomb excitation, followed by dissociation.  The
dissociation also provides a convenient experimental trigger.  Pair
production accompanied by mutual Coulomb excitation should occur at
smaller $b$, and have larger higher-order corrections than for
unaccompanied pairs.

Previous measurements of $e^+e^-$ pair production were at much lower
energies \cite{vane,ceres}.  The cross sections, pair masses, angular
and $p_T$ distributions generally agreed with the leading-order QED
perturbative calculations.  These studies did not require that the
nuclei break up, and so covered a wide range of impact parameters.

This letter reports on electromagnetic production of $e^+e^-$ pairs
accompanied by Coulomb nuclear breakup in $\sqrt{s_{NN}}= 200$ GeV per
nucleon pair Au-Au collisions \cite{vladimir}, as is shown in Fig. 1.
An $e^+e^-$ pair is produced from two photons, while the nuclei
exchange additional, independent photons, which break up the nuclei.
We require that there be no hadronic interactions, which is roughly
equivalent to setting the minimum impact parameter $b_{min}$ at twice
the nuclear radius, $R_A$, {\it i.e.} about 13 fm.  The Coulomb
nuclear breakup requirement selects moderate impact parameter
collisions ($2R_A < b < \approx 30$ fm) \cite{STARrho,baltzus}.
Except for the common impact parameter, the mutual Coulomb
dissociation is independent of the $e^+e^-$ production
\cite{hencken95,factorize}.  The cross section is
\begin{equation}
\sigma(Au Au \rightarrow Au^* Au^* e^+e^-) = \int d^2b P_{ee}(b) P_{2EXC}(b)
\end{equation}
where $P_{ee}(b)$ and $P_{2EXC}(b)$ are the probabilities of $e^+e^-$
production and mutual excitation, respectively at impact parameter
$b$.  The decay of the excited nucleus usually involves neutron
emission.  $P_{2EXC}(b)$ is based on experimental studies of neutron
emission in photodissociation \cite{newbaltz}.  For small $b$, a
leading-order calculation of $P_{2EXC}(b)$ may exceed 1.  A
unitarization procedure is used to correct $P_{2EXC}(b)$ to account
for multiple interactions \cite{baltzus,newbaltz}.

The most common excitation is a giant dipole resonance (GDR).  GDRs
usually decay by single neutron emission.  Other resonances decay to
final states with higher neutron multiplicities.  In mutual Coulomb
dissociation, each nucleus emits a photon which dissociates the other
nucleus.  The neutrons are a distinctive signature for nuclear
breakup.

\begin{figure}[tb]
\center{\includegraphics[width=0.8\columnwidth]{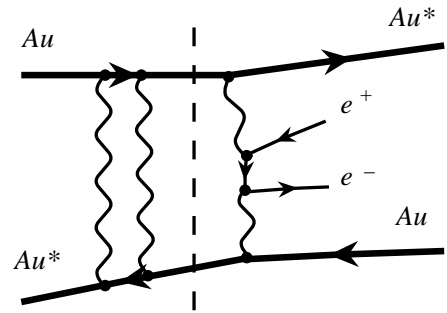}}
\caption[]{
\label{fig:diagram}
Schematic QED lowest-order diagram for $e^+e^-$ production accompanied
by mutual Coulomb excitation.  The dashed line shows the factorization
into mutual Coulomb excitation and $e^+e^-$ production.}
\end{figure}

We consider two different pair production calculations for
$P_{ee}(b)$.  The first uses the equivalent photon approach (EPA)
\cite{reviews}, which is commonly used to study photoproduction.  The
photon flux from each nucleus is calculated using the
Weizs\"acker-Williams method.  The photons are treated as if they were
real \cite{baurlum}.  The $e^+e^-$ pair production is then calculated
using the lowest-order diagram \cite{budnev}.  The photon $p_T$
spectrum for a photon with energy $k$ is given by
\cite{vidovic,interfere}
\begin{equation}
{dN\over dp_T} \approx  {F^2(k^2/\gamma^2 + p_T^2) p_T^2 \over
\pi^2 (k^2/\gamma^2 + p_T^2)^2}
\end{equation}
where $F$ is the nuclear form factor and $\gamma$ is the Lorentz boost
of a nucleus in the laboratory frame.  This calculation uses a
Woods-Saxon distribution with a gold radius of 6.38 fm and a 0.535 fm
skin thickness \cite{usPRC}.  The individual photon $p_T$ are added in
quadrature to give the pair $p_T$.  This is a minor simplification,
but should have little effect on the result.  For $e^+e^-$ pairs
visible in STAR, the typical photon $p_T\approx 3$ MeV/c, for a pair
$p_T\approx 5$ MeV/c.

The second calculation is a lowest-order quantum electrodynamics (QED)
calculation for pair production \cite{alscher}.  The main difference
between this calculation and the EPA approach is that the QED
calculation includes photon virtuality. In the relevant kinematic
range, the results of the calculations differ mainly in the pair $p_T$
spectrum \cite{kainew}.  In the QED calculation, the pair $p_T$ is
peaked at 20 MeV/c, higher than with the EPA.

One unavoidable difficulty in studying this reaction at an ion
collider is that $e^+e^-$ pairs are dominantly produced with a
forward-backward topology.  The angle between the electron momentum
and the two-photon axis in the two-photon rest frame, $\theta^*$, is
usually small.  Only a small fraction of the pairs are visible in a
central detector, limiting the statistics.

This analysis presents data taken in 2001 with the Solenoidal Tracker
at RHIC (STAR) detector at the Relativistic Heavy Ion Collider (RHIC).
Tracks were reconstructed in a large cylindrical time projection
chamber (TPC) \cite{TPC} embedded in a solenoidal magnetic field.
The track position and specific energy loss ($dE/dx$)
were measured at 45 points at radii between 60 and 189 cm from the
collision point.  Many of the tracks used in this analysis had low
transverse momenta, $p_T$, and curved strongly in the magnetic field,
and therefore had less than 45 reconstructable points.
This analysis used data taken in a 0.25 T magnetic field (half the
usual value).  

This analysis used about 800,000 events selected by a minimum bias
trigger \cite{trigger}.  This trigger selected events where both gold
nuclei broke up, by detecting events with one or more neutrons in zero
degree calorimeters (ZDCs) \cite{ZDCs} upstream and downstream of the
collision point.  The two ZDC hits were required to be within $1$
nsec of each other.  With the beam conditions and ZDC resolution, this
selected events along the beam line within $\approx$ 30 cm of the
detector center.

The signature for $e^+e^-$ production is two reconstructed tracks
which formed a primary vertex along the beamline and which had
specific energy losses consistent with those of electrons.  Event
vertices were found by an iterative procedure \cite{vladimir}.  The
analysis accepted events with a vertex containing exactly 2 tracks.
Up to two additional non-vertex tracks were allowed in the event, to
account for random backgrounds.

Tracks were required to have $p_T>65$ MeV/c and pseudorapidity
$|\eta|< 1.15$.  In this region, the tracking efficiency was above
80\%.  Tracks were also required to have momenta $p<130$ MeV/c, where
$dE/dx$ allowed good electron/hadron separation.  In this region, the
identification efficiency was almost 100\%, with minimal
contamination.  Pairs were required to have masses 140 MeV $< M_{ee} <
$ 265 MeV. The pair mass spectrum falls steeply with increasing
$M_{ee}$, so few leptons from pairs were expected with higher momenta.
Pairs were required to have $p_T< 100$ MeV/c and rapidity
$|Y|<1.15$. The pair cuts remove a very few background events, but
leave the signal intact.  These cuts selected a 
sample of 52 events.

\begin{figure*}[tb]
\center{
\includegraphics[width=0.95\columnwidth,clip]{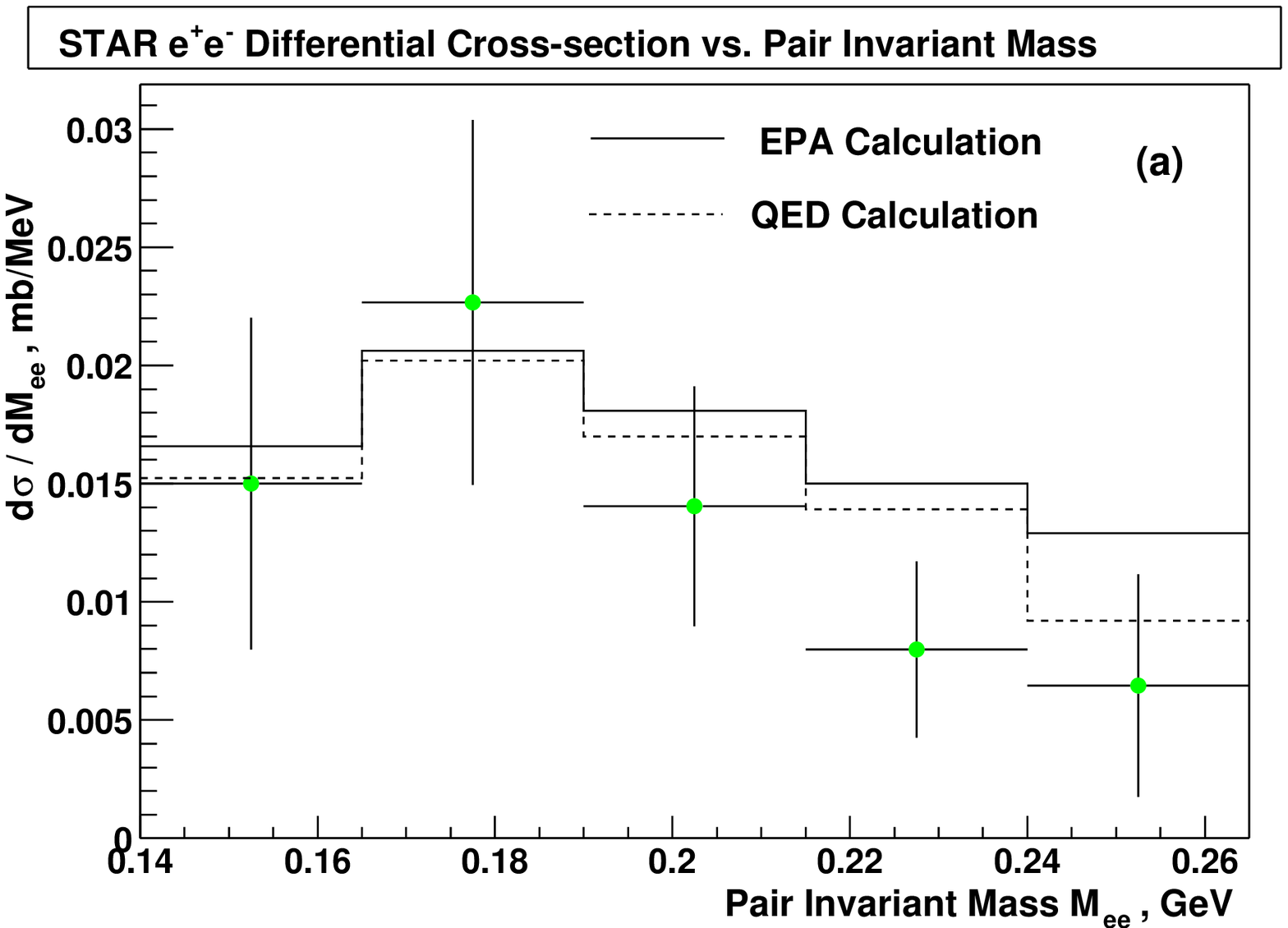}
\includegraphics[width=0.95\columnwidth,clip]{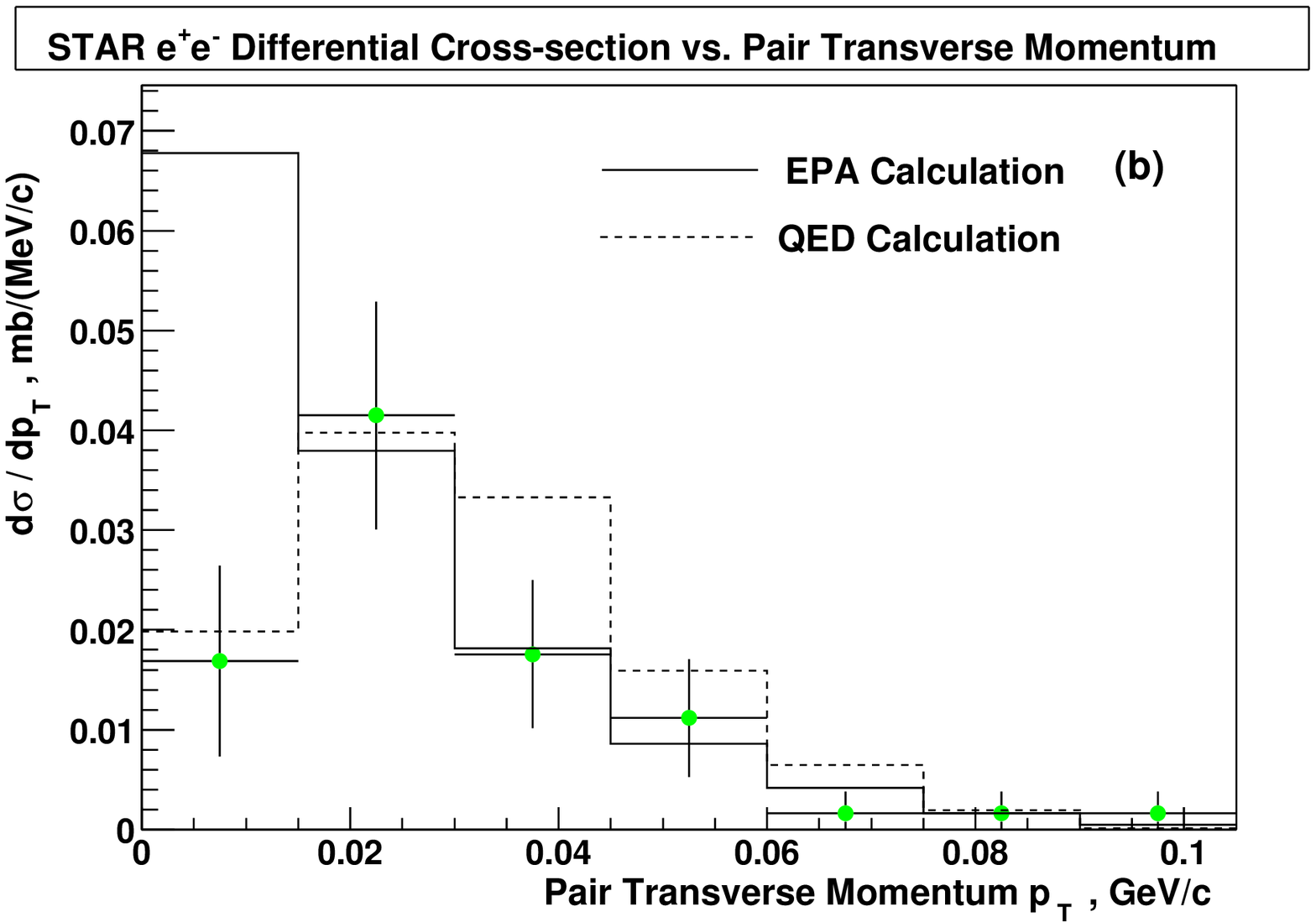}
}
\vskip .01 in
\center{
\includegraphics[width=0.95\columnwidth,clip]{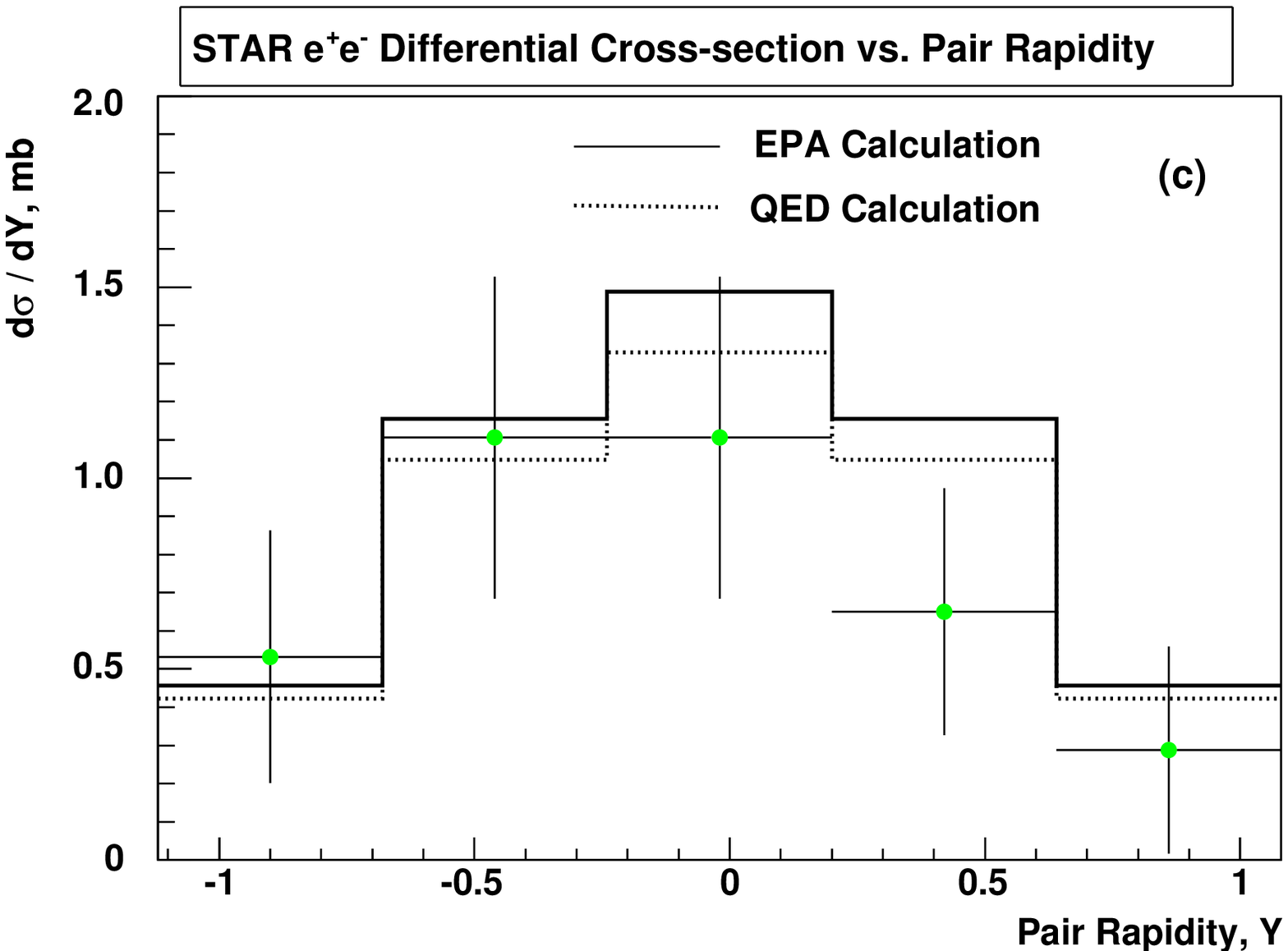}
\includegraphics[width=0.95\columnwidth,clip]{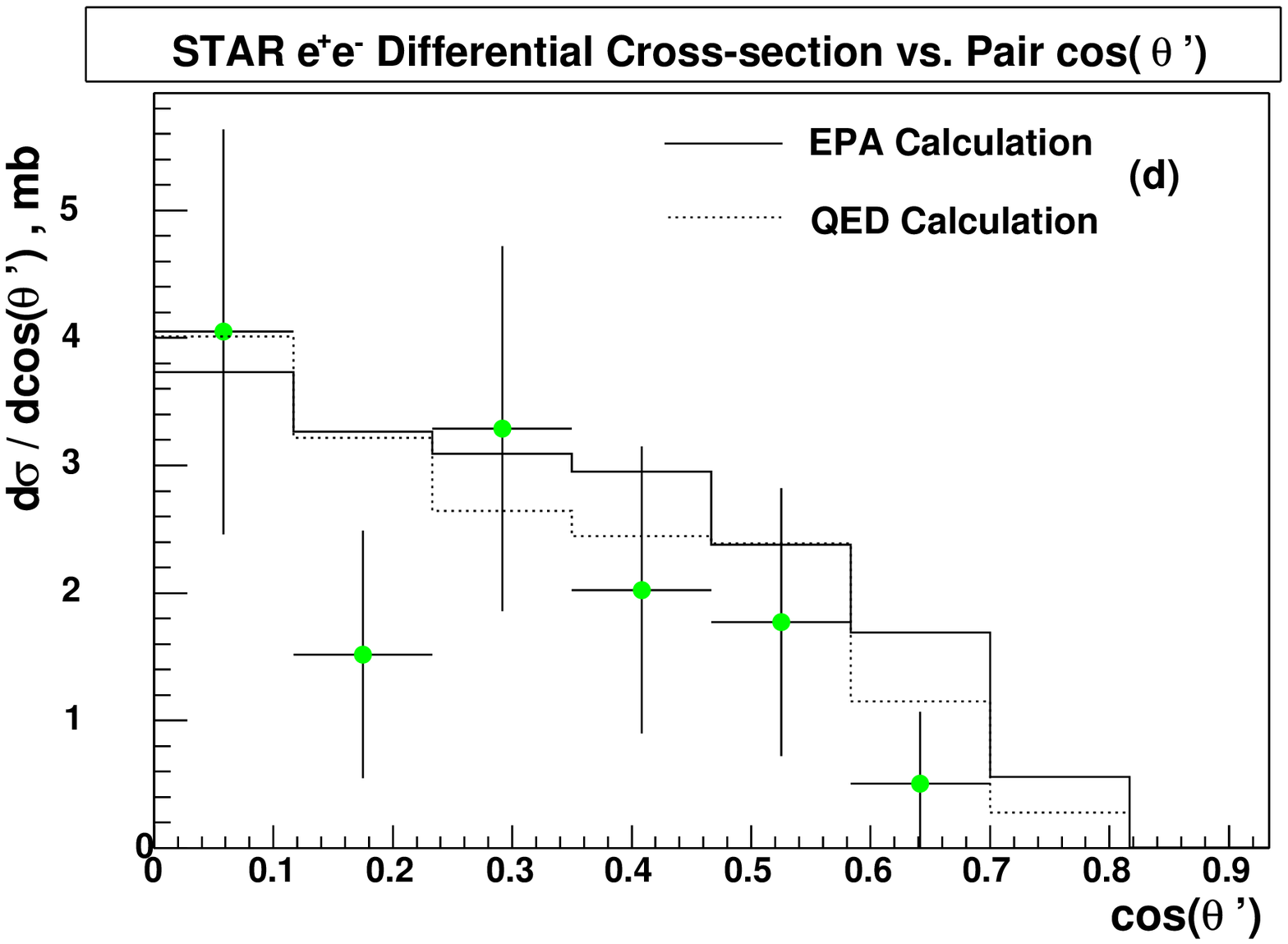}
}
\caption[]{
\label{fig:all}
(a) The pair mass distribution, (b) pair $p_T$, (c) pair rapidity and
(d) pair cos($\theta '$) distributions.  The data (points) are
compared with predictions from the EPA (solid histogram) and
lowest-order QED (dashed histogram) calculations.  The error bars
include both statistical and systematic errors.}
\end{figure*}

The data were corrected for efficiency using simulated events based on
the equivalent photon calculation and the standard STAR detector
simulation and reconstruction programs. The distributions of the
number of hits and track fit quality, the vertex radial positions and
track distance of closest approach matched in the data and simulations
\cite{vladimir}.

The resolutions were found to be 0.017 for pair rapidity, 0.01 for
track rapidity and 6 MeV for pair mass.  The pair $p_T$ resolution
varied slightly with $p_T$, but averaged about 4 MeV/c.  After
accounting for this $p_T$ smearing, the efficiency was found to be
independent of $p_T$.

There are two backgrounds in this analysis.  Incoherent (mostly
hadronic) backgrounds produce both like-sign and unlike-sign pairs, at
a wide range of $p_T$.  Based on a  study of like-sign and of higher
$p_T$ pairs, we estimate that this background is 1 event.  Coherent
backgrounds are due to photoproduction of mis-identified $\pi^+\pi^-$
pairs on one of the nuclei.  This background is peaked at higher
$M_{ee}$ than real $e^+e^-$ pairs.  
From the known $\rho^0$ and direct $\pi^+\pi^-$ cross sections
\cite{STARrho,usPRC}, and electron mis-identification probabilities,
the contamination is estimated to be less than 0.1 events. 
Backgrounds from other electromagnetic processes should be even
smaller.  The background from cosmic rays is suppressed to a
negligible level by the ZDC coincidence requirement.

The luminosity was determined by counting hadronic interactions with
at least 8 charged tracks.  This criteria selects 80\% of all hadronic
gold-gold interactions \cite{lum1,vladimir}.  After compensating for
the different neutron multiplicities in the hadronic and $e^+e^-$
samples (the ZDC timing resolution depends on the number of neutrons)
and assuming a total hadronic cross section of 7.2 barns
\cite{STARrho}, we find a total integrated luminosity of $94 \pm 9$
mb$^{-1}$.

The major systematic errors were due to uncertainties in the tracking
efficiency (6.4\% per track, or 13\% total), vertexing (8.5\%) and in
the luminosity (10\%) \cite{vladimir}.  The uncertainties due to
backgrounds and particle identification were much smaller and are
neglected.  These uncertainties were added in quadrature, giving a
18.5\% total systematic uncertainty.

Figure~\ref{fig:all}a shows the cross section for Au Au$\rightarrow$
Au$^*$Au$^*$ $e^+e^-$ as a function of pair mass, within our kinematic
fiducial region: track $p_T>65$ MeV/c, track pseudorapidity
$|\eta|<1.15$, pair rapidity $|Y|<1.15$ and pair mass 140 MeV $<
M_{ee}<265$ MeV.  The data are compared to the equivalent photon
(solid) and QED (dotted) calculations.  Monte Carlo events were
generated using the two calculations, and then filtered to match the
acceptances used here.  Both calculations match the pair mass data.

Figure~\ref{fig:all}b shows the cross section as a function of pair
$p_T$.  The equivalent photon (solid) and QED (dashed) calculations
differ when $p_T < 15$ MeV/c, due to the non-zero photon virtuality in
the QED calculation.  The data agree with the QED calculation, but not
with the equivalent photon calculation.

Figure \ref{fig:all}c shows the cross section as a function of pair
rapidity.  The broad peak around $Y=0$ is due to the detector
acceptance.  Selecting tracks with pseudorapidities $|\eta|<1.15$
preferentially chooses events with small pair rapidity.
The data agrees with both calculations. 

Figure \ref{fig:all}d shows the angular distribution $\cos(\theta')$
between the $e^+$ momentum and the beam axis, in the pair rest
frame. There is a small (usually $<5$ mrad) difference between
$\theta'$ and $\theta^*$ since the photon $p_T$ rotates the
$\gamma\gamma$ rest frame slightly with respect to the beam axis.  The
distribution in Fig. \ref{fig:all}d is the convolution of the detector
acceptance (largest at small $\cos(\theta')$) with the production
distribution, which is peaked at large $\cos(\theta')$.  The agreement
between the data and the calculations is good.

Within the kinematic range 140 MeV $< M_{ee} < 265$ MeV, pair rapidity
$|Y|<1.15$, track $p_T>65$ MeV/c and $|\eta|<1.15$, the cross
section $\sigma = 1.6\pm 0.2\ {\rm (stat)} \pm 0.3\ {\rm (syst)} $
mb, in reasonable agreement with the equivalent photon
prediction of 2.1 mb and the QED calculation of $\sigma_{QED} = 1.9$
mb.  At a 90\% confidence level, higher order corrections to the cross
section, $\Delta\sigma = \sigma-\sigma_{QED}$, must be within the
range $-0.5\sigma_{QED} < \Delta\sigma < 0.2\sigma_{QED}$.

\begin{figure}[tb]
\includegraphics[width=0.95\columnwidth,clip]{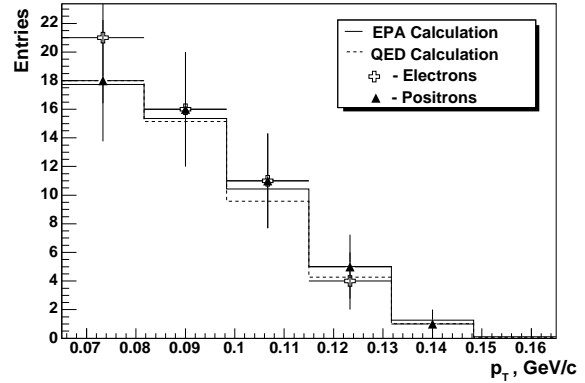}
\caption[]{The $p_T$ spectra of the produced electrons and positrons,
along with the comparable EPA and QED calculations.  In both
calculations, the electron and positron spectra are identical. Spectra
from the two calculations are similar; the data agrees with both of
them.}
\label{trackpt}
\end{figure}

At leading-order, the electron and positron momentum spectra are
identical.  However, interference with higher order corrections can
create charge-dependent spectral differences \cite{lee2}. For some
kinematic variables, 30-60\% asymmetries may occur \cite{brodsky}.  A
study of $e^+e^-$ production in sulfur-nucleus collisions at
$\sqrt{s_{NN}}=20$ GeV per nucleon pair found that the positrons had a
higher average energy than the electrons \cite{vane}.  However, the
atomic electrons in the target could have contributed to the result.
Figure \ref{trackpt} compares the $p_T$ spectra of the produced
electron and positron; the two spectra are very similar.  No asymmetry
is seen beyond the experimental uncertainties.

In addition, we have measured the fraction of events with a single
neutron in each ZDC to be $0.06\pm 0.04$ (3 out of 52).  This is
consistent with the single neutron fraction observed in similarly
tagged $\rho$ photoproduction \cite{STARrho}, supporting the notion of
independence assumed in the factorization, Eq. (1).

In conclusion, we have observed $e^+e^-$ production accompanied by
nuclear excitation in gold on gold ion collisions at a center of mass
energy of 200 GeV per nucleon pair.  The cross section, pair mass and
angular and rapidity distributions are in agreement with two
calculations, one using equivalent photons, and the other a
lowest-order QED calculation. The pair $p_T$ spectrum agrees with the
QED calculation, but not the equivalent photon calculation.  
Lowest-order QED describes our data.  We set a limit on
higher-order corrections to the cross section, $-0.5\sigma_{QED} <
\Delta\sigma < 0.2\sigma_{QED}$ at a 90\% confidence level.  The
electron and positron $p_T$ spectra are similar, with no evidence of
higher order corrections due to interference.

We thank Kai Hencken for providing the results of his QED calculation,
and Joakim Nystrand and Anthony Baltz for the nuclear breakup
subroutines used in the EPA calculation.  We thank the RHIC Operations
Group and RCF at BNL, and the NERSC Center at LBNL for their
support. This work was supported in part by the HENP Divisions of the
Office of Science of the U.S.  DOE; the U.S. NSF; the BMBF of Germany;
IN2P3, RA, RPL, and EMN of France; EPSRC of the United Kingdom; FAPESP
of Brazil; the Russian Ministry of Science and Technology; the
Ministry of Education and the NNSFC of China; SFOM of the Czech
Republic, FOM and UU of the Netherlands, DAE, DST, and CSIR of the
Government of India; the Swiss NSF.

\def\etal{{\it et al.}}

\end{document}